\documentclass[12pt]{iopart}

\usepackage{graphicx}
\newcommand{\lSAW}{\lambda_\mathrm{SAW}}	
\newcommand{\kSAW}{k_\mathrm{SAW}}			
\newcommand{\vSAW}{v_\mathrm{SAW}}			
\newcommand{\fSAW}{f_\mathrm{SAW}}			

\newcommand{\Prf}{P_\mathrm{rf}} 				
\newcommand{\Pl}{P_\ell}						
\newcommand{\mycomment}[1]{}						

\newcommand{\textsubscript}[1]{$_{#1}$}

\def\dEg{\delta E_\mathrm{g}}

\begin{document}

\title{Dynamics of indirect exciton transport by moving acoustic fields}

\author{A. Violante}
\address{Paul-Drude-Institut f\"ur Festk\"orkperelektronik, Hausvogteiplatz 5-7, 10117 Berlin, Germany}
\ead{violante@pdi-berlin.de}
\author{K. Cohen}
\address{Racah Institute of Physics, The Hebrew University of Jerusalem,  Jerusalem 91904, Israel}
\author{S. Lazi$\acute{\mathrm{c}}$}
\address{Paul-Drude-Institut f\"ur Festk\"orkperelektronik, Hausvogteiplatz 5-7, 10117 Berlin, Germany}
\address{Departamento de F\'isica de Materiales, Universidad Aut\'onoma de Madrid, E-28049 Madrid, Spain}
\author{R. Hey}
\address{Paul-Drude-Institut f\"ur Festk\"orkperelektronik, Hausvogteiplatz 5-7, 10117 Berlin, Germany}
\author{R. Rapaport}
\address{Racah Institute of Physics, The Hebrew University of Jerusalem,  Jerusalem 91904, Israel}
\author{P. V. Santos}
\address{Paul-Drude-Institut f\"ur Festk\"orkperelektronik, Hausvogteiplatz 5-7, 10117 Berlin, Germany}

\begin{abstract}
We report on the modulation of  indirect excitons (\textit{IX}s) as well as their transport by moving periodic potentials produced by surface acoustic waves (SAWs). The  potential modulation induced by the SAW strain modifies both the band gap and the electrostatic field in the quantum wells confining the \textit{IX}, leading to changes in their energy. In addition, this potential capture and transports \textit{IX}s over several hundreds of $\mu$m. While the \textit{IX} packets 
keep to a great extent their spatial shape during transport by the moving potential, 
the effective transport velocity is lower than the SAW group velocity and increases 
with the SAW amplitude. This behavior is attributed to the capture of \textit{IX}s by traps along the transport 
path, thereby increasing the \textit{IX} transit time. 
The experimental results are well-reproduced by an analytical model for the interaction between trapping centers and IXs during transport.

\end{abstract}

\maketitle

\section{Introduction}

The strong interaction with photons makes exciton ideal particles for the processing of optical information using solid-state devices.~\cite{Warburton_NP4_676_08} To reach this goal, functionalities like storage, transport, and manipulation of excitons become mandatory. Spatially indirect (or dipolar) excitons (\textit{IX}s) are particularly suitable for these applications due to their long lifetime and energy tunability. An \textit{IX} is a bound state of an electron and a hole localized in an undoped double quantum well (DQW) structure separated by a thin (i.e., less than the exciton Bohr radius) barrier. As illustrated in Figs.~\ref{Fig1}(a) and \ref{Fig1}(b), \textit{IX}s are formed by applying an electric field $F_z$ across the DQWs to drive the electron and hole constituents into different quantum wells (QWs), while still maintaining the Coulomb correlation between the particles. $F_z$ controls, via the quantum confined Stark effect (QCSE), both the energy and the lifetime of the \textit{IX}s, the latter of which can reach several hundreds of $\mu$s. \cite{Sivalertporn_PRB85_45207_12} 
These long lifetimes open the way for information storage as well as for the creation of cold bosonic gases for study of coherent exciton phases~\cite{Butov_N418_751_02,High_Nature_12,Shilo_a1305.2895v1__13}. 

The oriented \textit{IX} electric dipoles  give rise to strong repulsive \textit{IX}-\textit{IX} interactions.\cite{PVS239} The non-linearity associated with these interactions  has been explored for the realization of \textit{IX} gates~\cite{PVS239,Baldo_NP3_558_09} and exciton transistors~\cite{High_S321_229_08}.  The combination of non-linearity with tight lateral confinement using electrostatic gates has recently been explored for the isolation of single \textit{IX} states.\cite{Schinner_PRL110_127403_13} 
Also, the electric dipole provides a tool for the manipulation of \textit{IX}s using electric field gradients. Examples of electrically driven device functionalities include exciton storage cells, switches, and transistors~\cite{Winbow_NL7_1349_07,High_S321_229_08,Grosso_NP3_577_09}.

Long \textit{IX} lifetimes also enables the long-range transport of  \textit{IX} packets, which opens the way for the coupling of remote \textit{IX} systems. Different approaches have also been introduced to transport of  \textit{IX}s  based on bare diffusion,\cite{Voros_PRL94_226401_05} drift induced by repulsive \textit{IX-IX} interactions,\cite{Butov_N418_751_02} or by using spatially varying electric fields, including electrostatic ramps~\cite{Gaertner_APL89_052108_06} as well as moving electrostatic lattices.~\cite{Winbow_PRL106_196806_11}  Recently, we have shown that \textit{IX}s can be efficiently transported by the moving (and tunable) type-I band gap modulation ($\delta E_g$) produced by the strain of a surface acoustic wave [SAW, cf. Fig.~\ref{Fig1}(c)].~\cite{PVS177} During the SAW-induced transport, the long-living \textit{IX}s are captured in regions of minimum band gap, which move with the well-defined SAW velocity. The transport of  \textit{IX} packets with a constant velocity becomes interesting since it allows synchronization with control gates, optical sources, and detectors.\cite{Winbow_NL7_1349_07} 

The band structure modulation as well as the  long-range acoustic transport of electrons and holes by piezoelectric SAWs are well documented  in the literature.\cite{Rocke97a,PVS107,PVS109,PVS140} In this case, the interaction between the SAW and carriers is primarily mediated by the SAW piezoelectric field, which creates a moving type-II modulation of the conduction (CB) and valence band (CB) edges. This type of modulation separates electrons and holes and leads to exciton dissociation. The acoustic manipulation of excitons rather requires a type-I band gap modulation, which can be produced by the strain of a non-piezoelectric SAW.\cite{PVS177}  
While the piezoelectric modulation amplitudes can reach several tens of meV, the type I modulations is limited to a few meV. 
Due to the very different amplitudes, the  two types of modulations lead  to different transport regimes. 
Investigations of the  \textit{IX} dynamics under non-piezoelectric SAWs  have  so far only been carried out in the quasi-static regime, i.e., for time scales much longer than the acoustic period. 

In this work, we report on the  modulation of the energy levels  as well as on the transport of  \textit{IX} packets  in (Al,Ga)As quantum well structures 
by non-piezoelectric SAWs. 
The studies were carried out using spatially and time-resolved photoluminescence (PL) spectroscopy to probe the energetics as well as the evolution of \textit{IX} packets in space and time with resolutions in the $\mu$m and ns ranges, respectively. 
We show that the application of a SAW induces shifts in the \textit{IX} energies with very different time scales. 
The first are the oscillating changes in band gap caused by the strain field, which take place at the SAW frequency. The second  have a higher amplitude and  persist for much longer (up to ms) time scales. They are attributed to 
charge trapping in the (Al,Ga)As structure, which modifies the external electric fields applied to create IXs. 
In addition, the moving SAW fields  can transport  \textit{IX} packets over several hundreds of $\mu$m with reduced shape distortion. 
The transport velocity of the \textit{IX} packets  increases with acoustic amplitude but never reaches the SAW velocity,  even when  strong acoustic fields are applied. This behavior is attributed to trapping centers along the transport path, which capture \textit{IX}s and increase their transit time. We present a model for the effects of exciton trapping, which well reproduces  their impact on both the transport velocity and the shape of the \textit{IX} packets.

In the following, we first describe the sample structure and the optical techniques used to probe the acoustic transport of excitons by SAWs (Sec.~\ref{Experimental_Details}). We then present experimental optical studies of the effects of the acoustic fields on the \textit{IX} energy levels as well as on the time evolution of \textit{IX} packets during acoustic transport (Sec.~\ref{Results}). In Sec.~\ref{Discussion} the experimental results of \textit{IX} dynamics are analyzed in terms of a model for the transport, which takes into account \textit{IX} trapping by centers along the transport path. Finally, Sec.~\ref{Conclusions} summarizes the main conclusions of this study.


\section{Experimental details}
\label{Experimental_Details}

\begin{figure}[tbhp]
\begin{center}
\includegraphics[height=12cm, keepaspectratio=true,angle=0,clip]{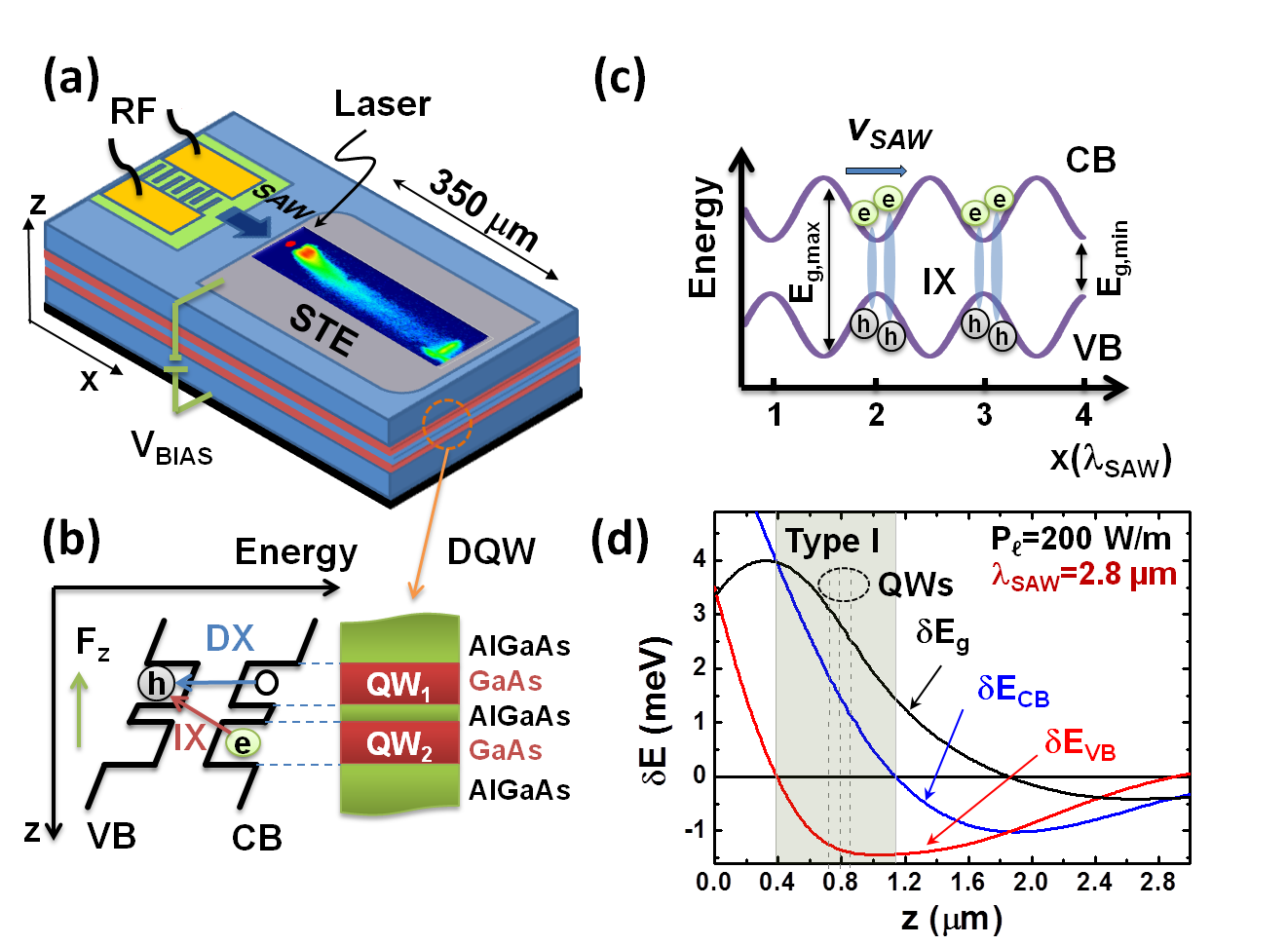}
\end{center}
\caption{(a) Samples for indirect exciton (\textit{IX}) transport by surface acoustic waves (SAWs). The \textit{IX}s are optically excited using a focused laser pulse in a double quantum well (DQW) structure by applying a bias $V_\mathrm{BIAS}$ between the doped substrate and a semitransparent top electrode (STE). The SAW is launched along the $x=[100]$ direction by an interdigital transducer (IDT) placed on a piezoelectric ZnO island. \textit{IX} transport is mapped by photoexciting \textit{IX}s in the DQW  and imaging the PL emitted along the transport path. The superimposed PL image was recorded at 3.8~K using $V_\mathrm{BIAS}=-1.6$~V 
and a band gap modulation amplitude $\delta E_g=1.8$~meV. (b) Energy band diagram of the GaAs/(Al,Ga)As DQW along the vertical ($z$) direction showing the direct (\textit{DX}) and \textit{IX} transitions under the electric field $F_z$. 
(c) \textit{IX} transport by the moving modulation of the conduction (CB) 
and valence bands (VB) in a DQW.  $E_\mathrm{g,min}$  and $E_\mathrm{g,max}$ 
denote, respectively, the minimum and maximum band gaps induced by the SAW strain field along the $x=[100]$ direction. 
(d) Depth dependence of amplitude modulation of the conduction ($\delta E_{CB}$) and valence  band edges ($\delta E_{VB}$) as well as of the bandgap modulation $\delta E_g=(\delta E_\mathrm{CB}-E_\mathrm{VB})$ calculated for a SAW with wavelength $\lSAW=2.8~\mu$m and linear power density $\Pl=200$~W/m. 
The band edge modulation is of type-I within the shaded region. The depth of the DQWs (dashed vertical lines) was selected to ensure that $|\delta E_\mathrm{CB}|\approx |\delta E_\mathrm{VB}|$.}
\label{Fig1}
\end{figure}

The samples used in the experiment consist of three sets of asymmetric GaAs DQWs grown by molecular beam epitaxy on a $n^{+}$-doped GaAs(001) substrate [cf.~Fig.~\ref{Fig1}(a)]. Each DQW is formed by  a 14~nm (QW$_1$) and a 17~nm-thick QW (QW$_2$) separated by a thin (4~nm-thick) Al\textsubscript{0.3}Ga\textsubscript{0.7}As barrier. Multiple DQWs were used in order to enhance the photon yield in the optical experiments.  The different QW widths allows for the selective generation  of excitons in either QW$_1$ or QW$_2$ by the appropriate selection of the excitation wavelength:\cite{Stern_PRL101_257402_08} this feature, however, has not been  explored in the present studies.  The electric field $F_{z}$ for the formation of \textit{IX}s was induced by a bias voltage $V_\mathrm{BIAS}$ applied  between the n-doped substrate and a thin (10 nm-thick) semitransparent Ti electrode (STE) deposited on the sample surface, as illustrated in Fig.~\ref{Fig1}(b).

The moving acoustic field was provided  by a Rayleigh SAW with a wavelength $\lSAW$=2.8 $\mu$m corresponding to a frequency $\fSAW=1$~GHz at 4~K. The SAW was generated along a $\langle100\rangle$ non-piezoelectric surface direction using aluminum interdigital transducers (IDTs) deposited onto a piezoelectric ZnO island. In order to minimize the screening of the  radio-frequency (rf) field by carriers in the doped substrates, the DQW stack were  embedded within thick,  undoped Al$_{0.26}$Ga$_{0.74}$As layers. In this way, the total thickness of the undoped overlayers could be increased to approx. one $\lSAW$.  

The use of non-piezoelectric SAWs is essential to prevent exciton ionization by the piezoelectric field.\cite{PVS177}  As illustrated in Fig.~\ref{Fig1}(c), the strain field produces, via the deformation potential interaction, a moving type-I modulation of the band edges, which confines \textit{IX}s close to the SAW phases of lower band gap  [cf.~Fig.~\ref{Fig1}(c)] and transports  them with the acoustic velocity $\vSAW$.
The SAW strain field and, therefore, the modulation of the band edges change with depth. The solid lines in Fig.~\ref{Fig1}(d)  display the depth dependence of the conduction  ($\delta E_{CB}$) and valence  ($\delta E_{VB}$) bands, as well as as the band gap modulation $\delta E_g=(\delta E_\mathrm{CB}-E_\mathrm{VB})$ calculated for a SAW with a linear power density  $\Pl=200$~W/m ($\Pl$ is defined as the acoustic power per unit length perpendicular of the SAW beam).
The profiles  were determined using a $k\cdot p$ approach by taking into account the deformation potentials and the full SAW strain field obtained from an elastic continuum model for SAW propagation in the layered structure of the sample.\cite{PVS156} 
The amplitude and relative phases of the band edged modulation changes with depth. For depths  between 0.15~$\lSAW$ and 0.35~$\lSAW$ $\delta E_{CB}$ and $\delta E_{VB}$ have opposite phases, thus producing a type-I lateral potential modulation, which confines  electrons and holes at the same spatial position $x$ along the SAW path. The dashed vertical lines represent the positions of  the three DQWs, which were selected in order to get band edge modulations of approximately the same amplitude but opposite phases for electrons and holes.\cite{PVS177} Different depths also lead to slightly different modulation amplitudes $\delta E_g$  in the three DQWs. For the range of used acoustic powers, however, these  differences are small compared to the spectral linewidth of the IXs.

The spectroscopic experiments were carried out at temperatures between 1.8 and 3.8~K in a bath He cryostat with rf electrical connections for SAW excitation. Excitons were generated on the SAW path using a laser beam focused onto a spot with a diameter of 10~$\mu$m using a microscope objective. We used either a tunable  cw Ti:Sapphire laser or a pulsed semiconductor laser emitting at 780~nm with  pulse duration and repetition rate  of 300~ps and 2.5~MHz, respectively. 
The laser excitation energies were always below the band gap of the (Al,Ga)As barrier in Fig.~\ref{Fig1}(b)  to selectively excite carriers in the QWs.
 The PL from \textit{IX}s  emitted along the SAW path was collected by the same objective and detected with energy and spatial resolution using a cooled charge-coupled-device (CCD) camera connected to a spectrometer. Time-resolved PL measurements were carried out using an avalanche photodiode with time resolution of 400~ps synchronized with the laser pulses.

     
%

\section{Results} 
\label{Results}

\subsection{Bias dependence of the \textit{IX} lines}               

\begin{figure}[tbhp]
\begin{center}
\includegraphics[height=10cm, keepaspectratio=true,angle=0,clip]{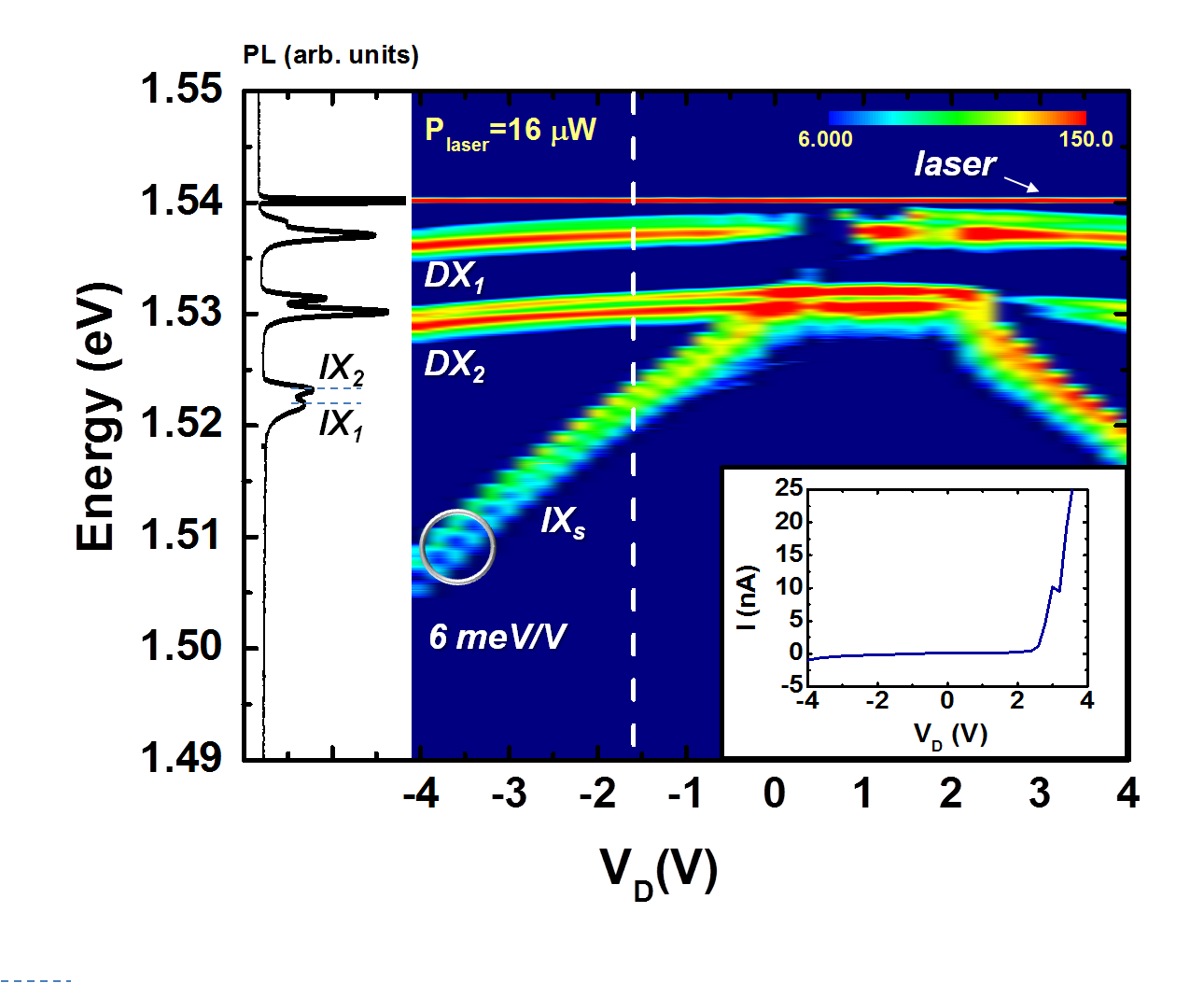}
\caption{Dependence of the photoluminescence intensity emission (color scale) on energy and applied bias $V_\mathrm{BIAS}$ (cf.~Fig.~\ref{Fig1}(a)). The measurement were carried out at 2 K using an  excitation energy of 1.54~eV. $DX_1$ and $DX_2$ denote the emission lines associated with the electron-heavy hole direct excitons of QW$_1$ and QW$_2$, respectively.  The indirect exciton line ($IX$) red-shifts with reverse bias at a rate of 6~meV/V. The inset shows the I-V characteristic of the Schottky diodes, which have an area of approx.~$400\times400~\mu$m$^2$.  The right plot displays the spectrum measured at the voltage indicated by the dashed line.}
\label{FigPLVD}
\end{center}
\end{figure}

The structure of Fig.~\ref{Fig1}(b) forms a rectifying Schottky diode with the current ($I$) \textit{vs.} voltage ($V_\mathrm{BIAS}$) characteristics illustrated in the inset of Fig.~\ref{FigPLVD}. The diodes have an excellent blocking behavior in the reverse bias region (i.e., for negative $V_\mathrm{BIAS}$). The current levels for voltages in the range $-2<V_\mathrm{BIAS}<1$~V were less than 10~nA for devices with an area of $400\times400~\mu$m$^2$, even under excitation with light energies below the band gap of the (Al,Ga)As barriers. As will become clear latter,  these blocking characteristics are important to avoid carrier injection from the contacts and, thus, to ensure that both the optically generated electrons and holes are carried along the DQW structures  during acoustic transport.

The dependence of the PL emission lines on  $V_\mathrm{BIAS}$ is displayed in Fig.~\ref{FigPLVD}.  Here, $DX_1$ and $DX_2$ denote the emission from direct excitons associated with the electron-heavy hole transitions in $QW_1$ and $QW_2$, respectively. A closer observation (cf. left inset) reveals that the DX lines are split into two components.  The splitting is attributed to the coexistence of both neutral and charged direct excitons. The  $DX_1$ and $DX_2$ energies slightly red-shift with $|V_\mathrm{BIAS}|$ due to the intra-well QCSE induced by the applied  bias.

The application of a negative bias creates \textit{IXs} with the electron  and hole constituents confined within the wider (QW$_2$) and narrower (QW$_1$) quantum wells, respectively. 
In general, the  \textit{IX} PL is split in different lines (as indicated by $IX_1$ and $IX_2$ in Fig.~\ref{FigPLVD}), whose energy  red-shifts linearly with the applied bias.  The splitting between the \textit{IX} lines depends on acoustic power and on the  illumination conditions. 
This last behavior contrast with the one observed for direct excitons, where, for each QW,   the energy splitting between charged and neutral exciton lines does not change with the excitation conditions.  As will be discussed in detail later, the appearance of more than one \textit{IX} line is attributed to inhomogeneities in the electric field within the DQW region. The latter  can give rise to up to three different  \textit{IX}  energies, each corresponding to one of the DQWs.  
The \textit{IX} lines  red-shifts due to the inter-well QCSE at a rate of approximately $r_V=6$~meV/V. The latter corresponds approximately to the potential difference between the center plane of the QWs given by $V_\mathrm{BIAS}(d_{QW_1}+d_{QW_2}+d_B)/d_T=6.92$ meV/V, where $d_{QW_i}$ ($i=1,2$) denotes the QW thickness, $d_B$ is the barrier width, and $d_T$ the total thickness of the intrinsic regions between the doped substrate and the STE. Finally, formation of \textit{IX} is also observed for positive bias higher than approximately 2.5~V. This regime, however, will not be further explored here.

\subsection{Acoustic modulation of excitons}               
\label{Acoustic_modulation_of_excitons}

\begin{figure}[tbhp]
\begin{center}
\includegraphics[width=12cm, keepaspectratio=true,angle=0,clip]{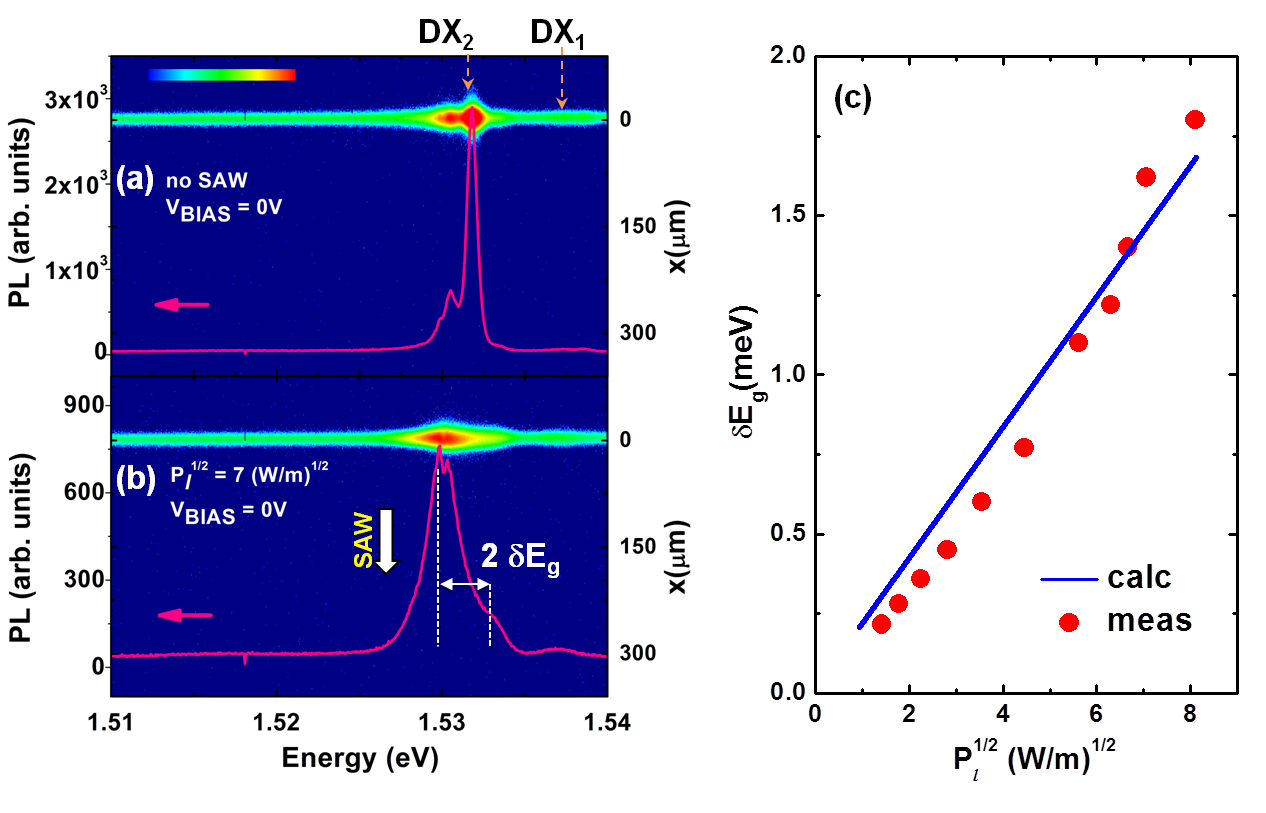}
\end{center}
\caption{Spectral PL maps (on a log scale)   as a function of the distance $x$ from the laser excitation spot recorded for nominal rf-powers applied the IDT of  (a) $\Prf=0$ and (b) $\Prf=12$~dBm. The measurements were carried out at 1.8~K under $V_\mathrm{BIAS}=0$~V, where only the electron-heavy hole direct excitons in QW$_1$ and QW$_2$ are observed ($DX_1$ and $DX_2$, respectively).
(c) Band gap modulation $\dEg=(E_\mathrm{g,max}-E_\mathrm{g,min})/2$ as a function of $\Pl^{1/2}$, where $\Pl$ is the acoustic power per unit length perpendicular of the SAW beam. The solid line was calculated following the $k\cdot p$ procedure described in the text.}
\label{BandGap}
\end{figure}

The color maps of Figs.~\ref{BandGap}(a) and (b) illustrate the effects of the acoustic field on the emission of DXs.  The plots  display the spectral distribution of the PL from the DQWs as a function of the distance ($x$) from the generation spot  in the absence and presence of a SAW beam, respectively. The measurements were carried out at 3.8~K with $V_\mathrm{BIAS}=0$~V. Under these conditions,  \textit{IX} do not form and the spectra are dominated by  direct  excitons  from QW$_1$ and QW$_2$. 
The DX emission is spatially constrained to a circle  of radius of approx. 16~$\mu$m around the generation spot. The size of the emission region   is  determined by expansion of the DX cloud around the illumination spot. In contrast to the behavior of excitons under piezoelectric SAWs,\cite{Rocke97a} neither a substantial quenching of the integrated PL intensity  nor  long-range acoustic transport are observed under non-piezoelectric SAWs. This behavior is consistent with the fact  that the type-I band gap  modulation of Fig.~\ref{Fig1}(b)  does not ionize excitons by  spatially separating electrons and holes.  The DX radiative lifetimes remain short, thus preventing the acoustic transport over distances $>>\lSAW$.

As in the piezoelectric case,~\cite{PVS107}  Fig.~\ref{BandGap}(b) also shows that the periodic modulation of the band gap by the SAW strain splits the DX \cite{PVS107,PVS177} lines,  the splitting being equal to twice the amplitude  $\delta E_\mathrm{g} = (E_\mathrm{g,max} - E_\mathrm{g,min})/2$ of the band gap modulation.
 $\delta E_\mathrm{g}$   is plotted as a function of $\Pl^{1/2}$ in Fig.~\ref{BandGap}(c). $\Pl^{1/2}$ has been chosen as the horizontal coordinate since it is proportional to the amplitude of the strain field induced by the SAW. $\delta E_\mathrm{g}$ reaches values up to 1.8~meV for the highest applied acoustic power. The solid line displays  the dependence of $\delta E_\mathrm{g}$ on $\Pl$ calculated using the previously mentioned $k\cdot p$ approach.
Both the measured and calculated values reproduce the linear dependence of the modulation amplitude on $\Pl^{1/2}$.

\begin{figure}[tbhp]
\begin{center}
\includegraphics[height=10cm, keepaspectratio=true,angle=0,clip]{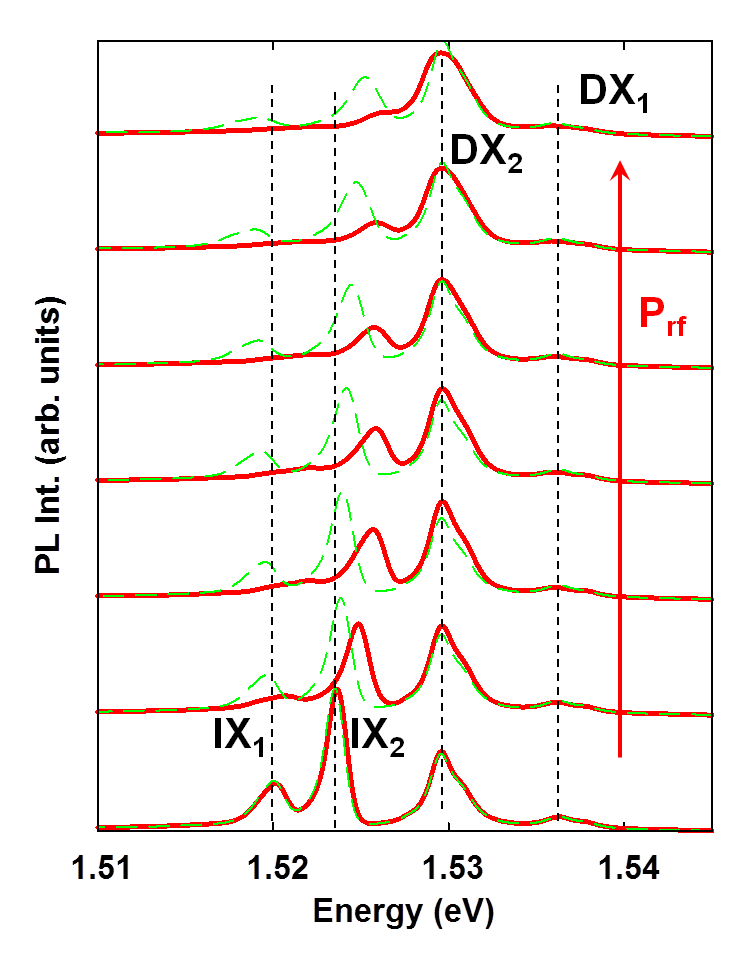}
\end{center}
\caption{Dependence of the PL spectrum on acoustic power (red solid lines). The measurements were performed at 2~K under $V_\mathrm{BIAS}=-1.6$~V for increasing powers $\Prf$ applied to the IDT (from bottom to top: no power, -4, 0, 4, 8, 12, and 16 dBm).  The red solid curves were recorded under the phase ON conditions  described in Sec.~\ref{Experimental_Details}. The dashed curves show, for comparison, the corresponding spectra measured in the OFF phase. The vertical dashed lines mark the energy of the DX and IX transitions in absence of acoustic excitation.}
\label{131207-020aInt}
\end{figure}

The impact of the acoustic fields on the PL spectrum is illustrated in Fig.~\ref{131207-020aInt}. The bottom spectrum  shows the PL spectrum of the samples  recorded under a reverse bias ($V_\mathrm{BIAS}=-1.6~V$) in the absence of a SAW. Under these conditions, one finds in addition to the direct exciton lines the red-shifted signatures from  \textit{IXs}. The additional solid curves in this figure were obtained for increasing acoustic powers (from bottom to top). The \textit{IX} lines broaden and shift in energy with $\Prf$. The broadening is attributed to the strain-induced modulation of the (spatially indirect) band gap. In contrast to the behavior of DXs under a small applied bias, the larger spectral width and longer lifetime of the \textit{IX}s prevents the observation of the strain-induced  splitting of the lines. 

The \textit{IX} energy shifts with increasing $\Prf$ in Fig.~\ref{131207-020aInt} are different for  lines $IX_1$ and $IX_2$ and much larger than the band gap modulation amplitude $\delta E_g$.  Note that while the center position of the DX emission lines remain constant, the $IX_2$ line blue-shifts and approaches DX$_2$ for high acoustic powers. This behaviour is attributed to changes in the electric field across the DQWs due to charge redistribution induced by the acoustic field. The DQWs are embedded within (Al,Ga)As barrier layers with high electrical resistance. As a results,  charges can be stored close to them, leading to variations in the electrostatic field configuration. We have observed, for instance, that illumination with energy above the band gap of the barrier layers can lead to large shifts in the \textit{IX} energies in the absence of acoustic excitation. Here, the electric field becomes perturbed by the photo-excited electrons and holes, which are spatially separated by the applied vertical field. For selective excitation of the QWs using laser energies below the band gap of the barriers, in contrast, these shifts become very small. However, they increase when a SAW is turned on (cf. Fig.~\ref{131207-020aInt}).

The SAW induced \textit{IX} energy shifts  in Fig.~\ref{131207-020aInt} have a transient character and  persists  after the acoustic field has been switched off.  In order to justify this assignment the PL was recorded while modulating both the amplitude of the rf-power $\Prf$ applied to the IDT and the laser excitation beam with a square wave with a repetition period of 3.6~ms. The solid lines in Fig.~\ref{131207-020aInt} were recorded with the optical and rf square waves in phase (denoted as the phase ON condition). In this case, both excitations are applied simultaneously to the sample.  For comparison, measurements were also carried out by shifting the phase of the two square waves  by 180$^\circ$ (phase OFF condition, indicated by the dashed lines in Fig.~\ref{131207-020aInt}). Here, the sample is illuminated while the SAW is switched off. Since the lifetime of the carriers is much shorter than the square wave period, the PL is emitted during the light pulses, when the SAW is turned off. Note, however, that with increasing $\Prf$ the \textit{IX} energies also blue-shift with respect to the energies measured without acoustic excitation (lowest spectrum), the blue-shift increasing with the amplitude of the previously applied SAW pulses. The observed  shifts in \textit{IX} energy indicate that the charge configuration created by the  SAW pulses  persists for times scales exceeding one ms. In contrast, the DX lines remains essentially the same in the phases ON and OFF.  Finally, since the modulation periods are also shorter than the thermal relaxation times, these results also show that the modifications in the \textit{IX} emission cannot be assigned to thermal effects.


The acoustically induced energy  shifts appear for very weak acoustic powers, i.e., for levels  far below those  required for long-range acoustic transport along the SAW propagation direction (see Sec.~\ref{Indirect_exciton _Transport}). 
In contrast to the changes due to illumination with energies above the barrier band-gap, the acoustically induced modifications  are probably associated by  carrier transport along the QW plane followed by charge trapping. 
The determination of the  microscopic mechanism responsible for this behavior requires further investigations and is beyond the scope of the present paper.

\subsection{Indirect exciton transport}               
\label{Indirect_exciton _Transport}

The optical technique used to probe acoustic transport is illustrated by the time-integrated PL image superimposed on the sample layout of Fig.~\ref{Fig1}(a), which maps the  average distribution of transported \textit{IX} along the channel  underneath the STE defined by the SAW beam.  
In this image, the contributions from exciton diffusion to the transport, which are only significant close to the generation spot, were eliminated by subtracting a similar  image recorded in the absence of the SAW beam. 
The PL  along the transport path is attributed to the recombination of \textit{IX}s captured by trapping centers in the DQW plane. While it allows to visualize the \textit{IX} distribution, trapping also reduces the transport efficiency and, as will be discussed in detail below,  the effective \textit{IX} transport velocity. The mechanism for the strong PL at the edge of the STE opposite to the IDT has a different origin. Here,  the potential barrier created by the abrupt reduction of the vertical field $F_z$ blocks further transport and forces \textit{IX} recombination.

\begin{figure}[htbp]
\begin{center}
 \includegraphics[height=8 cm, keepaspectratio=true,angle=0,clip]{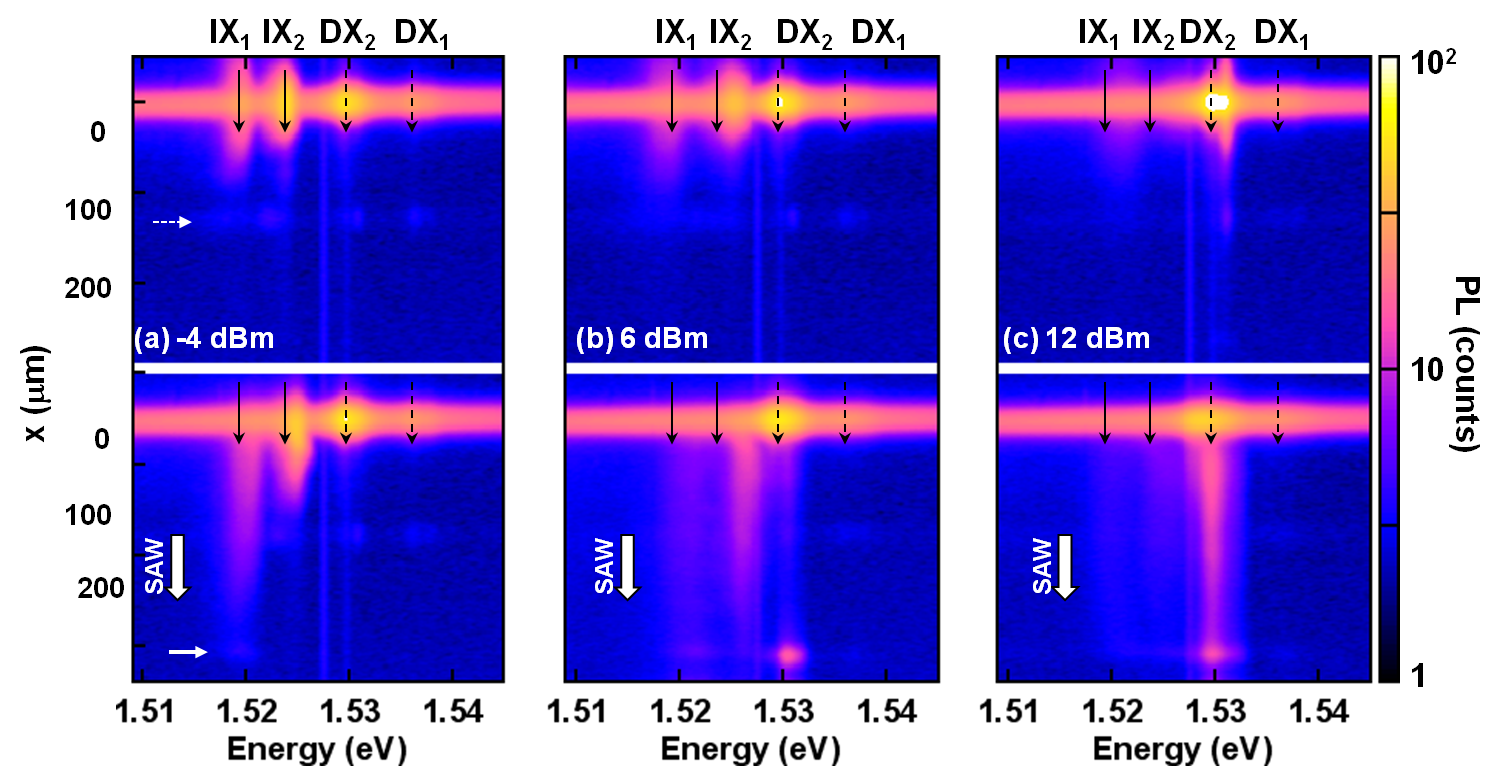}
\end{center}
\caption{ Spectral PL maps (on a log scale) along the SAW propagation direction  as a function of the distance ($x$) from the laser excitation spot recorded in the absence (upper panels, phase OFF) and presence of SAWs (phase ON) for  rf-powers ($\Prf$) of  (a) -4~dBm, (b) 6~dBm, and (c) 12~dBm.  The maps were recorded at 2~K using $V_\mathrm{BIAS}=-1.6$~V. The dashed and solid vertical arrows mark the positions of direct and indirect excitons, respectively, in the absence of acoustic excitation. 
The strong recombination at $x=250~\mu$m (horizontal solid arrow) coincides with the edges of the STE (cf. Fig.~\ref{Fig1}(a)). The emission at $x=150~\mu$m (horizontal dashed line in (a)) is due to recombination at a defect on the SAW path. The thin vertical lines at 1.528 and 1.530~eV result from stray light from a gas lamp.
}
\label{131207-020aN}
\end{figure}

Additional information about the \textit{IX} transport mechanism was obtained from the spectral distribution of the PL along the transport path measured for different  acoustic powers  in the lower panels of Fig.~\ref{131207-020aN}. These PL maps were recorded under the phase ON condition described above, where the optical and acoustic excitations are applied in phase. The \textit{IX} are captured by the acoustic field and transported up to the end of the STE, which is located at a distance $d=250~\mu$m away from the generation spot [indicated by the horizontal solid arrow in (a)]. The upper plots display, for comparison, the corresponding maps measured in the phase OFF conditions. The \textit{IX} cloud extends, in this case, up to distances from the generation point of at most $80~\mu$m. 

The solid and dashed vertical arrows in Fig.~\ref{131207-020aN} mark the energy of the main indirect and direct exciton lines determined in the absence of acoustic excitation. For low acoustic power [lower panel in Fig.~\ref{131207-020aN}(a)] transport takes place preferentially through the lowest lying $IX_1$ state, which has  the longest recombination lifetime. For intermediate SAW amplitudes  [Fig.~\ref{131207-020aN}(b)], the strongest emission along the transport path is observed for $IX_2$, which  blue-shifts with acoustic power. In contrast,  most of the PL  at the end of the transport channel has an energy close to the one of   DX$_2$. Since the lifetime of direct excitons is expected to be much  smaller than the transit time of the carriers (on the order of $d/\vSAW=100$~ns, where $\vSAW$ is the acoustic velocity), they cannot be  transported over such large distances. The emission at the DX$_2$  energy along the transport path is attributed to the conversion of \textit{IX}s to DX$_2$, followed by a fast radiative decay of the DX$_2$s. The interconversion becomes enhanced at the edges of the STE, where the suppression of $F_z$ reduces the energy separation between  the direct and indirect excitonic species.

Under the action of a reverse bias $V_\mathrm{BIAS}$, the \textit{IX}s in the structure of Fig.~\ref{Fig1}(b) consist of electrons stored in the wider (QW$_2$) and holes in the narrower (QW$_1$) quantum wells. As a result, the IX-DX$_2$ conversion requires the excitation of the IX-holes from $QW_1$ to the wider $QW_2$. The efficiency of this process is expected to increase when the difference in energy between the hole levels in the two QWs becomes comparable to the amplitude of the acoustic modulation of the VB edge, on the order of 1-2~meV in the present case.   The latter is  supported by the results for high acoustic power in Fig.~\ref{131207-020aN}(c), when $IX_2$  approaches $DX_2$. In this case, most of the emission along the transport path takes place at the direct exciton energy. 

\subsection{Transport dynamics}

\begin{figure}[htbp]
\begin{center}
\includegraphics[height=12 cm, keepaspectratio=true, angle=0,clip]{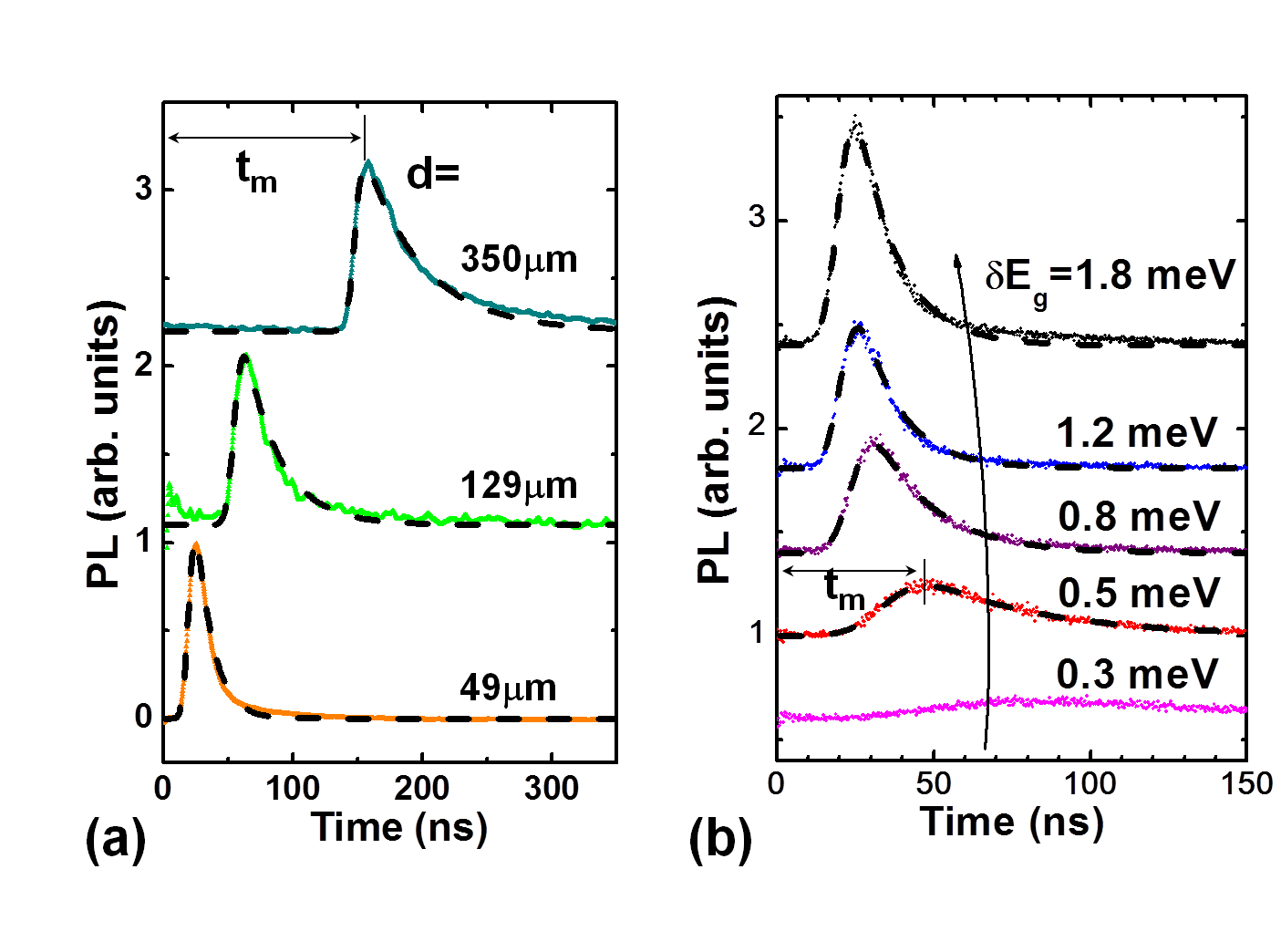}
\end{center}
\caption{
 (a) Time-resolved PL profiles from \textit{IX}s recorded at different distances $d$ from the laser spot under a SAW-induced band gap modulation  $\dEg=1.8$~meV. 
(b) Profiles recorded at $d=49~\mu$m for different $\dEg$. The dashed lines are fits to Eq.~(\ref{Eq2}).
}
\label{Fig2}
\end{figure}

 The dynamics of the acoustic transport was investigated by recording time-dependent PL profiles 
at different distances \textit{d} from the pulsed laser excitation spot using an avalanche photodiode synchronized with the laser pulses.  The integrated PL over the energy range of excitonic emission was collected over a 10~$\mu$m-wide stripe across the  propagation path of a SAW.   The time resolution of the experiments is then 
given by the ratio between this length and the average velocity of the \textit{IX} packets (see below).  The SAWs were generated by a continuous rf-source.
Figure~\ref{Fig2}(a) displays time-resolved profiles recorded for different transport distances   \textit{d} 
 under a strong acoustic modulation amplitude ($\delta E_\mathrm{g}=1.8$~meV). 
In all cases, well-defined PL pulses are observed with time delays $t_m$  increasing with propagation distance according to  $t_m=d/v_{IX}$, where $v_\mathrm{IX}$ is the average \textit{IX} transport velocity (cf.~Fig.~\ref{Fig2}).
In addition, the rising edge of the time-resolved PL pulses becomes less abrupt for increasing $d$ while the trailing edge develops a tail indicating a distribution of arrival times at the detection position.

The initial radius  of the \textit{IX} cloud created by the laser was estimated  from the size of the PL emission region around the laser spot [cf.~Fig.~\ref{BandGap}(a)] to be  $w_{IX}=16~\mu$m. The cloud extends, therefore, over  several SAW wavelengths $\lSAW=2.8~\mu$m. 
For the shortest transport distance ($d=49~\mu$m), the width $t_w$ of the PL 
 pulses corresponds closely to the ratio $w_\mathrm{IX}/v_\mathrm{IX}$ 
 between the initial Gaussian width ($w_\mathrm{IX}$) of the \textit{IX} cloud  and the average \textit{IX} transport velocity ($v_\mathrm{IX}$).

The effects of the SAW amplitude on the \textit{IX} dynamics are illustrated in  Fig.~\ref{Fig2}(b). Here, the PL profiles were recorded at a fixed distance $d=49~\mu$m while varying $\Prf$ to change the band gap modulation amplitude $\dEg$. The shape of the pulses is almost the same for  large modulation amplitudes ($\delta E_\mathrm{g}>1.2$~meV), under which the \textit{IX}s are efficiently trapped and transported by the SAW fields. For lower amplitudes, in contrast,  $t_m$ increases substantially whereas  
the PL pulses broaden and develop a pronounced tail.

\begin{figure}[htbp]
\begin{center}
 \includegraphics[height=10 cm, keepaspectratio=true, angle=0,clip]{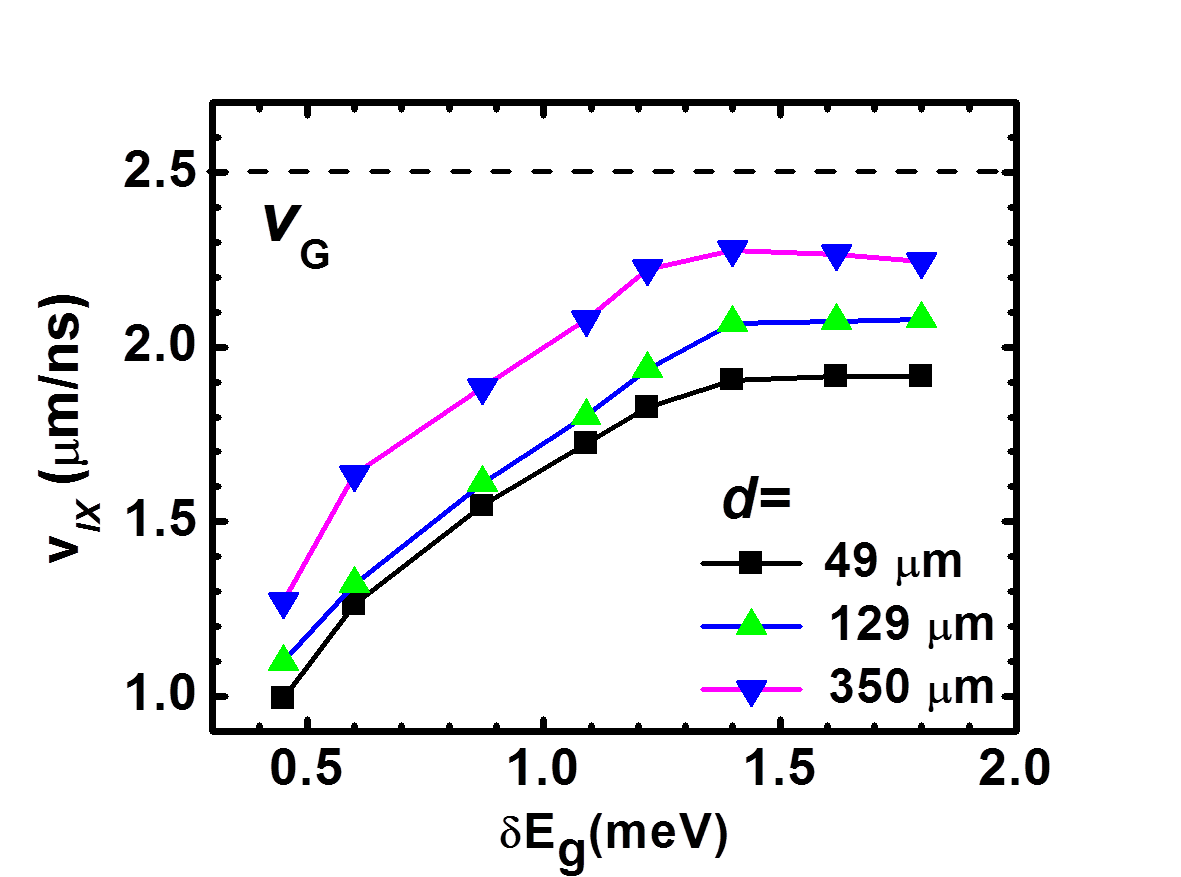}
\end{center}
\caption{Dependence of  the \textit{IX} velocity ($v_{IX}$) on the band gap modulation amplitude $\delta E_g$.  
}
\label{velEg}
\end{figure}

Figure~\ref{velEg} summarizes  the dependence of the average transport velocity $v_\mathrm{IX}$ on the band gap 
modulation amplitude ($\dEg$) recorded at different transport distances $d$. $v_\mathrm{IX}$ increases with $\dEg$ until it saturates for $\delta E_\mathrm{g}>1.2$~meV. Interestingly, the saturation velocity increases with $d$ and never reaches  the SAW group velocity along the transport path of $v_\mathrm{G}=(2.5\pm0.1)~\mu$m/ns (dashed line).
Due to the presence of the ZnO layer, $v_\mathrm{G}$  in the region in-between the ZnO islands differs from the product 
of the IDT resonance frequency and its wavelength.  $v_\mathrm{G}$ was determined 
by measuring the propagation time of an acoustic pulse in a delay line with two transducers 
built on the same sample. These measurements were carried out using a network analyzer with time-domain capabilities.

%
According to Figs.~\ref{Fig2}(b) and \ref{velEg},  long-range acoustic transport takes place for modulation amplitudes $\dEg>0.5$~meV, when the maximum effective exciton velocity in the acoustic potential,  $v_\mathrm{IX,max}$, exceeds the SAW velocity $v_\mathrm{G}$. 
 $v_\mathrm{IX,max}$ can be stated in terms of the exciton mobility $\mu_\mathrm{IX}$ as  $v_\mathrm{IX,max}=\mu_\mathrm{IX} \kSAW\dEg$, where $\kSAW=2\pi/\lSAW$. Using this expression, we estimate $\mu_\mathrm{IX}=v_\mathrm{IX}/( \kSAW\dEg)=2.2\times10^{4}$~cm$^2$/(eV s) and an \textit{IX} diffusion coefficient $D_\mathrm{IX}=7.3$~cm$^2$/s at 3.8~K. These values agree well with  those reported for \textit{IX}s in DQWs with comparable QW thicknesses.~\cite{Voros_PRL94_226401_05} Note, however, that the efficient transport of \textit{IX}s requires modulation amplitudes much larger (by a factor of approx. 3) than those for the onset of transport. 


\begin{figure}[htbp]
\begin{center}
\includegraphics[height=10 cm, keepaspectratio=true, angle=0,clip]{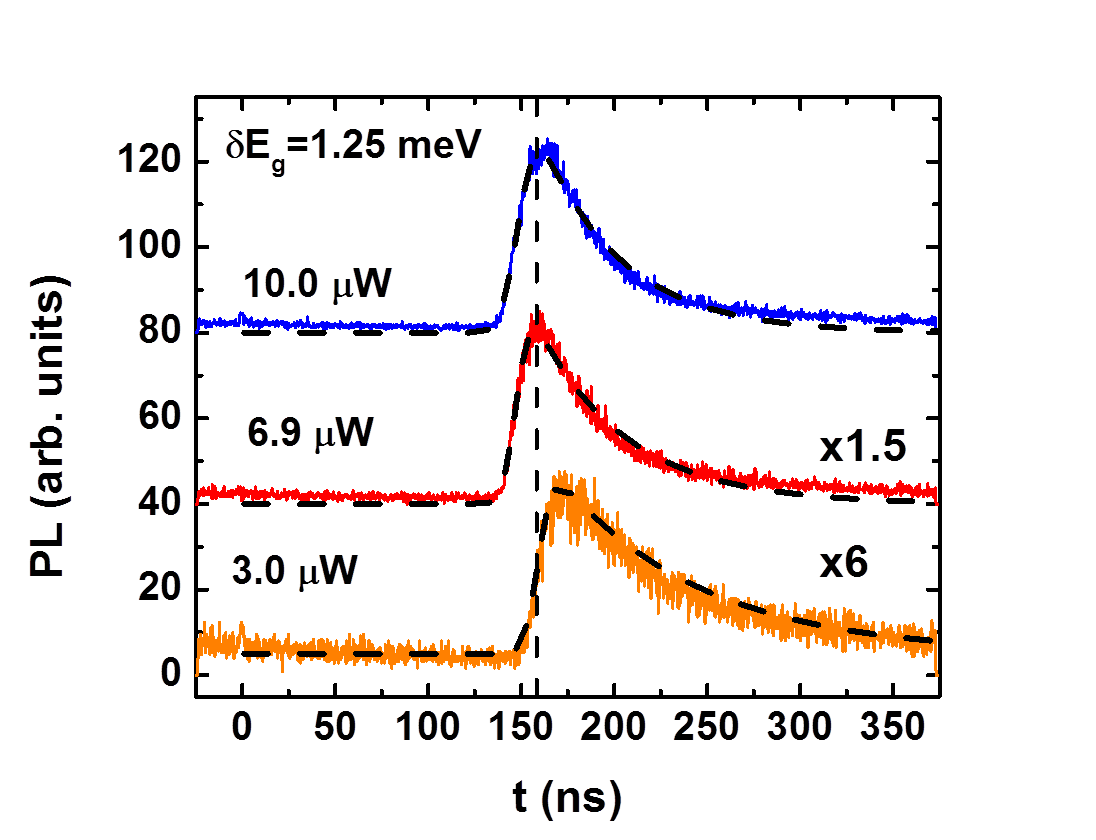}
\end{center}
\caption{
Time-resolved PL profiles from \textit{IX}s recorded for different light excitation fluxes for a  transport distance of 350~$\mu$m. The dashed lines are fits to the model described by Eq.~\ref{Eq2}. 
}
\label{IntDep}
\end{figure}

The dependence of the \textit{IX} dynamics on \textit{IX} density was investigated by recording time-resolved profiles as a function of the laser pulse intensity (cf. Fig.~\ref{IntDep}). 
The experiments were carried out in the regime of strong acoustic excitation (i.e., for $\delta E_g>1$~eV in Fig.~\ref{velEg}), where $v_{IX}$ is approximately independent of the SAW power. The profiles exhibit a weak dependence on the excitation density, thus indicating that non-linear effects are negligible in the range of studied densities.

\section{Discussions}
\label{Discussion}

Previous studies of acoustic transport efficiency using piezoelectric SAWs have identified two main regimes.\cite{PVS140} For small acoustic powers, the carriers are  dragged by the SAW with an effective  velocity much smaller than the acoustic velocity.  This behavior corresponds to dynamics of \textit{IX}s  observed for $\delta E_g\leq0.5~$~meV in Fig.~\ref{Fig2}(b).  In contrast,  high acoustic modulations efficiently trap the carriers, as depicted in  Fig.~\ref{Fig1}(c). In this regime,  a Gaussian \textit{IX} packet with initial width $t_w=w_{IX}/v_{IX}$ should move  with the SAW group velocity $v_\mathrm{G}$ while maintaining its shape. The time-dependent PL profiles at a distance \textit{d} would then be given by 
%
\begin{equation}
g(t,d)=A_g \exp\left[{-\left(\frac{t-t'_m}{\sqrt{2} t_w}\right)^2}\right],
\label{Eq1}
\end{equation}
%
\noindent  where $A_g$ is the maximum PL intensity and $t'_m=d/v_G$. The \textit{IX} kinetics displayed in Figs.~\ref{Fig2}(a) and \ref{velEg}, however, does not follow this prediction: the exciton velocity $v_{IX}$ increases with 
the propagation distance and never reaches the SAW group velocity $v_\mathrm{G}$. In addition, the \textit{IX} 
packets broaden (indicating an increase in $t_w$)  for increasing transport distances and develop a tail towards longer times.

\begin{figure}[htbp]
\begin{center}
 \includegraphics[height=7 cm, keepaspectratio=true, angle=0,clip]{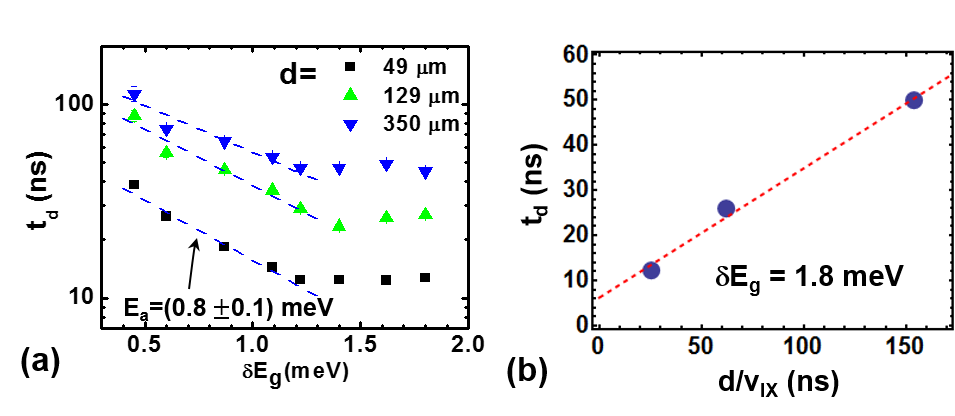}
\end{center}
\caption{
Dependence of   the characteristic trapping time ($t_d$)
on (a) the  band gap modulation amplitude $\delta E_g$ and on   (b) the  transport time $d/v_{IX}$ recored for high modulation amplitudes ($\delta E_g=1.8$~meV).
}
\label{model}
\end{figure}

We  mentioned in Sec.~\ref{Acoustic_modulation_of_excitons}  that carrier trapping  may change the potential distribution in the DQWs, thus leading to changes in  the energy of  the \textit{IX} species. We show in this section  that trapping 
 can also account for the reduced transport velocity as well as for the deviations of the profile shapes from Eq.~(\ref{Eq1}). 
In order to model the effects of the trapping centers, we will assume that they can capture and retain the \textit{IX}s over an interval $t$ described by the exponential distribution $f(t) = \frac{1}{t_d} e^{- \frac{t}{t_d}}$. Here, $t_d$ is the average time the \textit{IX}s remain immobilized in traps during  transport. In this approximation, an initially Gaussian profile [of the form in Eq.~(\ref{Eq1})] will  evolve in time according to
%
\begin{eqnarray}
\label{Eq2}
h(t,d) & \approx&  (g\ast f)(t,d)\\ \nonumber
&=& 
  \sqrt{\frac{\pi }{2}} 
  \frac{A_g}{t_d} e^{-\frac 1 2 \left(\frac{ t_w}{ t_d}\right)^2}
\mathrm{Erfc}\left(-\frac{t-{t_m}}
{\sqrt{2} {t_w}}\right)
   e^{-\frac{t-t_m'}{t_d}},
\end{eqnarray}
%
\noindent where $\mathrm{Erfc}(t)$ is the complementary error function,  $t_m = t'_m+\frac{t^2_w}{t_d}$, 
and  "$\ast$" denotes the convolution operator. The approximation given by Eq.~\ref{Eq2} is valid for $t_d>t_w$, which is almost always satisfied.  

While the maximum of Eq.~\ref {Eq1} occurs for  $t=t'_m$, the maximum of Eq.~(\ref {Eq2}) (as well as for the spectra of   Fig.~\ref {Fig2}) takes place for $t > t_m$.  
One of the main effects of trapping is to increase the transit time due to the extra delay  $t_w^2/t_d$, leading to  the reduced   \textit{IX} saturation velocity $v_{IX}=d/( t_m)<d/t'_m$  displayed in Fig.~\ref{velEg}. The extra delay is related to the width of the leading edges of the PL pulses described by $t_w$. While  $t_d$ increases with propagation distance $d$ (see discussion below), $t_w$ is determined by a diffusion-like process and increases proportionally to $d^\eta$ with $\eta\approx<1/2$. 
The increase of the saturation values of $v_{IX}$ with $d$ arises from the fact 
that the relative contribution of factor $\frac{t^2_w}{t_d}\propto d^{1/2}$ reduces with increasing $d$.


The predictions of the model are shown by the solid lines superimposed on the data points in Figs.~\ref{Fig2} and \ref{IntDep}, which  were obtained  by using  $t'_m$, $t_d$,  and  $A_g$ of Eq.~(\ref{Eq2}) as fit parameters. The  quality of the fits dependents weakly on $t_w$ except for the lowest modulation 
amplitudes ($\dEg\leq0.5$~meV). In these cases, the fits improve by varying also $t_w$ [as in Fig.~\ref{Fig2}(b)]. The excellent agreement with the experimental results over a wide range of  transport distances and modulation amplitudes is a strong indication that the model correctly describes the \textit{IX} dynamics  during acoustic transport. 


Further information about the trapping centers can be obtained from the characteristic trapping times $t_d$ determined by the fittings, which are summarized in Fig.~\ref{model}(a). The dependence of $t_d$  on $\dEg$ mimics the one displayed for $v_{IX}$ in Fig.~\ref{velEg}. In fact, $t_d$ decreases exponentially with $\dEg$ with a characteristic decay energy $E_a=(0.8\pm 0.1)$~meV in the region of low acoustic amplitudes and saturates for $\dEg>1.2$~meV.  $E_a$  is assigned to the energy required for acoustic  excitation of trapped \textit{IX}s to the transport path.
Note that it exceeds the thermal energy at the measurement temperature ($k_BT  = 0.33$~meV) by more than a factor of two.
 
The trapping time  $t_{d}$ saturates for high SAW amplitudes, thus indicating that the transport in this regime becomes controlled by trapping centers with capture and emission kinetics  that do not depend on the acoustic intensity.   Finally, $t_d$ also increases  with the transport time $d/v_{IX}$. This behavior is attributed to the increase in the number of 
trapping events with transport distance. In order to justify this assertion, we plot in Fig.~\ref{model}(b) the dependence of $t_d$ on the transport time $d/v_\mathrm{IX}$ in the saturation regime (for $\dEg>1.2$~meV). $t_d$ increases linearly with the transport time:  from the extrapolation to zero transport times, we obtain a characteristic dwell time of 6~ns for each trapping event.

\section{Conclusions}
\label{Conclusions}

We have carried out a detailed investigation of the effects of high-frequency acoustic fields on dipolar \textit{IX}s in (Al,Ga)As DQW structures. We have shown that the  SAW strain shifts in the excitonic energies via two mechanisms. The first is the modulation of the band gap by the SAW strain, which produces mobile potentials for the trapping and transport of IXs. The second arises from electrostatic changes in the electric field applied across the DQWs induced by acoustically induced charge trapping. While the former takes place in the nanosecond time scale defined by the SAW frequency, the latter are transient and persist over times exceeding one millisecond. We have also  carried out a detailed investigation of the \textit{IX} transport dynamics  using time-resolved techniques. These studies have shown that the  \textit{IX}s  are carried as well-defined packets moving  with velocities close to the SAW velocity.  The dynamics of the 
packets is determined by trapping centers along the transport path, which capture \textit{IX}s and reduce 
their effective transport velocity.  We present a detailed model  the transport process  including the effect of traps, which reproduces very well both the time and the spatial evolution of \textit{IX} packets during acoustic transport. 

The strong impact of traps on the \textit{IX} dynamics is partially associated with the small band gap modulation induced by the SAW strain, which is comparable to the \textit{IX} spectral linewidths. Different approaches can be followed to increase the transport efficiency. One of them consists in improving sample growth conditions to increase the mobility and reducing, e.g., the density of trapping sites. A second explores the enhancement of the acoustic fields. In fact, since the acoustic power densities produced by piezoelectric transducers scale with the $1/\lSAW^2$, much strong modulation amplitudes are expected for high frequency SAWs. Finally, the present experiments were carried out using relatively small exciton densities. For higher densities, one can also take advantage of the  repulsive interactions between \textit{IX}, which can smooth potential fluctuations~\cite{Remeika_PRL102_186803_09} and, therefore, increase the mobility.
\\
We thank Holger T. Grahn for discussions and comments on the manuscript. We are indebted to S. 
Rauwerdink, W. Seidel, S. Krau\ss, and A. Tahraoui for the assistance in the preparation 
of the samples. We also acknowledge financial support from the German DFG (grant 
SA-598/9).

\vspace{ 1 cm}


\def\litdir{y:/myfiles/jabref}


\end{document}